\documentclass[12pt,a4paper,final]{iopart}

\usepackage{braket}
\usepackage{graphicx}
\usepackage{dcolumn}
\usepackage{bm}
\usepackage{setspace}
\usepackage{amssymb}
\usepackage{color}

\newcommand{\Yb}{\ensuremath{^{171}\mathrm{Yb}^+~}}
\newcommand{\avg}[1]{\ensuremath{\left\langle#1\right\rangle}}

\begin{document}
\title{}{\bf \LARGE Verification of the Quantum Nonequilibrium Work Relation in the Presence of Decoherence}

\author{Andrew Smith$^{1*}$, Yao Lu$^{2*}$, Shuoming An$^{2}$, Xiang Zhang$^{2}$\footnote{Present address: Department of Physics, Renmin University, Beijing, China}, Jing-Ning Zhang$^{2}$, Zongping Gong$^{3}$, H. T. Quan$^{3,4\dagger}$, Christopher Jarzynski$^{5,6\dagger}$, Kihwan Kim$^{2\dagger}$}
\address{ $^1$ Department of Physics, University of Maryland,  College Park, MD 20742 USA \\ 
$^2$ Center for Quantum Information, Institute for Interdisciplinary Information Sciences, Tsinghua University, Beijing 100084, P. R. China \\
$^3$ School of Physics, Peking University, Beijing 100871, P. R. China\\ 
$^4$ Collaborative Innovation Center of Quantum Matter, Beijing 100871, P. R. China\\
$^5$ Institute for Physical Science and Technology, University of Maryland,  College Park, MD 20742 USA\\
$^6$ Department of Chemistry and Biochemistry, University of Maryland,  College Park, MD 20742 USA }
\eads{ \mailto{htquan@pku.edu.cn, cjarzyns@umd.edu, kimkihwan@mail.tsinghua.edu.cn}}

\begin{abstract}
Although nonequilibrium work and fluctuation relations have been studied in detail within classical statistical physics, extending these results to open quantum systems has proven to be conceptually difficult. For systems that undergo decoherence but not dissipation, we argue that it is natural to define quantum work exactly as for isolated quantum systems, using the two-point measurement protocol. Complementing previous theoretical analysis using quantum channels, we show that the nonequilibrium work relation remains valid in this situation, and we test this assertion experimentally using a system engineered from an optically trapped ion. Our experimental results reveal the work relation's validity over a variety of driving speeds, decoherence rates, and effective temperatures and represent the first confirmation of the work relation for non-unitary dynamics.
\end{abstract}
%
\section{Introduction}
Statements of the second law of thermodynamics are generally expressed as inequalities.
For instance the work performed on a system during an isothermal process must not exceed the net change in its free energy: $W \ge \Delta F$.
When statistical fluctuations are appropriately included these inequalities can be reformulated as equalities, such as the nonequilibrium work relation~\cite{Jarzynski1997a}
\begin{equation}
\label{EQ:JE}
\langle e^{-\beta W} \rangle = e^{-\beta \Delta F}
\end{equation}
where $\beta$ is an inverse temperature and angular brackets denote an average over repetitions of the process.
For classical systems, this prediction and related {\it fluctuation theorems} have been extensively studied both theoretically \cite{Jarzynski2011} and experimentally \cite{Liphardt2002, Collin2005, Douarche2005, Blickle2006, Harris2007, Junier2009, Shank2010, Saira2012}, and have been applied to the numerical estimation of free energy differences \cite{Chipot2007,Pohorille2010}.

The last decade has seen growing interest in extending these results to quantum systems \cite{Hanggi2015}.  This pursuit has been hindered by the fact that classical work is defined in terms of trajectories -- a notion that is typically absent in the quantum setting.  To avoid this complication, many studies have focused on closed quantum systems, which evolve unitarily.
In this situation there is no heat transfer to or from the system
and the first law of thermodynamics reads,
\begin{equation}
\label{eq:Wdef}
W = \Delta U \equiv E_f - E_i \quad .
\end{equation}
Here the classical work depends only on a system's initial and final configuration and can be determined from two measurements.   This idea is easily lifted to the quantum regime through the {\it two-point measurement} (TPM) protocol \cite{Kurchan2000, Tasaki2000, Mukamel2003}, according to which the work performed during a single experimental run is the difference between energy values $E_i$ and $E_f$ resulting from initial and final projective measurements.  Note that this approach to measuring work is valid only for initial system states which lack coherence in the energy basis as these states are undisturbed by the initial measurement~\cite{Kammerlander16}.  This restriction does not hinder the following as we will always consider systems which begin in equilibrium.

If a system is prepared in equilibrium at inverse temperature $\beta$ with initial Hamiltonian $\hat{H}(0) = \sum \epsilon_n | n \rangle \langle n |$, then evolves unitarily as the Hamiltonian is varied from $\hat{H}$(0) at $t=0$ to $\hat{H}(\tau) = \sum \bar{\epsilon}_m \vert \bar{m} \rangle \langle \bar{m} \vert$ at $t=\tau$, the TPM work distribution is given by
\begin{equation}
\label{EQ:WD}
p(W) = \sum_{nm} p_n \, p_{\bar m\vert n} \, \delta[W - (\bar{\epsilon}_m - \epsilon_n)].
\end{equation}
Here $p_n = Z_0^{-1}e^{-\beta\epsilon_n}$ is the probability to obtain the value $E_i = \epsilon_n$ during the initial energy measurement, $p_{\bar m\vert n}$ is the conditional probability to obtain the final energy value $E_f = \bar{\epsilon}_m$, given the initial value $\epsilon_n$, and $Z_0$ is the partition function for the initial equilibrium state.
To date, both proposed~\cite{Huber2008,Dorner2013,Mazzola2013,Paz2014} and implemented~\cite{Batalhao2014, An2015,2017arXiv170305885N,2017arXiv170502990M} experimental tests of the quantum work relation (Eq.~\ref{EQ:JE}) have focused on evaluating Eq.~\ref{EQ:WD} for a closed system.

A number of authors have proposed definitions of work and derived fluctuation theorems for quantum systems in contact with general thermal environments \cite{Yukawa2000,Maes2004,Esposito2006,Crooks2008,Campisi2009}.
Our more focused aim in this paper is to consider a quantum system in contact with a thermal environment that produces decoherence but no dissipation.
From a theoretical viewpoint, we argue that the TPM protocol provides a natural definition of quantum work in this situation, and we give an elementary, physically motivated derivation of Eq.~\ref{EQ:JE} that agrees with more general results obtained by previous authors \cite{Rastegin2013,Rastegin2014,Kafri2012,Albash2013,Jordan2015}.
We then describe an experimental implementation constructed from trapped ions that simulates an externally driven system subject to decoherence but no dissipation.
From the data we verify the validity of the quantum work relation, providing the first experimental confirmation of Eq.~\ref{EQ:JE} for a system undergoing decoherence.

\section{Theoretical Development}
When a quantum system is coupled to a thermal environment, there arise two distinct
departures from unitary dynamics: dissipation, that is the exchange of energy, and decoherence, the leakage of the system's quantum coherences into the environment~\cite{Zurek2002}.
\textit{We will consider situations in which dissipation is negligible over experimentally relevant time scales, but decoherence is substantial}.
Under such conditions the environment is a {\it decohering} (or {\it dephasing}) environment: it suppresses coherences but does not exchange energy.

Consider a system in contact with a decohering environment.
At $t=0$, following a projective energy measurement, the system begins in an energy eigenstate $| \epsilon_n \rangle$, then it evolves as its Hamiltonian is varied with time.
At $t=\tau$ its energy is again measured, yielding $\bar{\epsilon}_m$.
By assumption, no energy is exchanged with the environment, therefore we claim that it is natural to identify work to be the difference between the initial and final energies, $W = \bar{\epsilon}_m - \epsilon_n$, just as for a closed quantum system (Eq.~\ref{eq:Wdef}).
If we accept this as a plausible definition of work in the presence of a decohering environment, then does Eq.~\ref{EQ:JE} remain valid in this situation?
This question can be answered affirmatively within the general framework of quantum channels~\cite{Rastegin2013,Rastegin2014,Jordan2015}.
We now take a phenomenological approach to arrive at the same answer.

We begin by modeling the dynamics of the system.
In the energy representation, a decohering environment does not affect the diagonal elements (populations) of the system's density matrix $\hat\rho(t)$, but may cause off-diagonal matrix elements (coherences) to decay.
We capture these features with the equation
\begin{equation}
\label{DME}
\frac{d\hat\rho}{dt} = -\frac{i}{\hbar}[\hat{H}(t), \hat\rho ] - \sum_{i\neq j} \gamma_{ij} \rho_{ij}  |i\rangle \langle j | \,\,\equiv \,\, {\cal L} \hat\rho \quad,
\end{equation}
which describes both unitary evolution under $\hat H(t)$ and the decohering effects of the environment.
Here $\gamma_{ij} \ge 0$ are phenomenological decay rates for the coherences $\rho_{ij}\equiv \langle i \vert\hat\rho\vert j\rangle$, in the instantaneous eigenbasis of $\hat H(t)$.

Although we have motivated Eq.~\ref{DME} heuristically, it can also be obtained from the perspective of {\it quantum detailed balance master equations} (QDBME) \cite{Alicki1976}.  These equations are a special type of Lindblad master equation and are of physical relevance as they rigorously describe a quantum system coupled to an infinite, thermal quantum reservoir under appropriate assumptions of weak interaction and separation of time scales~\cite{Kossakowski1977,Spohn1978,Gorini1978}.

For an $N$-level quantum system with no degenerate energy gaps as shown in Appedix A, the QDBME governing the evolution of the density operator can be written in the form
\begin{eqnarray} \label{QDBME}
\frac{d \hat{\rho}}{dt} &=& -\frac{i}{\hbar}[\hat{H},\hat{\rho}] + \sum_{ij} J_{ij} |i\rangle \langle i | + \sum_{i\ne j} \Gamma_{ij}  |i\rangle \langle j | \\ \nonumber
J_{ij} &\equiv& R_{ij} \rho_{jj} - R_{ji} \rho_{ii} \\ \nonumber
\Gamma_{ij} &\equiv& (R_{ii} + R_{jj} - \gamma_{ij}) \rho_{ij} < 0 \\ \nonumber
\gamma_{ij} &\equiv& \sum_k d_k (O_{ki} - O_{kj})^2 \geq 0
\end{eqnarray}
where the $R_{ij}$'s form a stochastic rate matrix~\cite{VKampen2007} satisfying detailed balance, the $O_{ij}$'s form a real orthogonal matrix, and $d_k>0$ for all $k$.
The three terms on the right side of Eq.~\ref{QDBME} respectively describe unitary evolution, dissipation, and decoherence.
The dissipative term evolves the diagonal elements of $\hat{\rho}$ (populations) according to a classical Markov process described by the rate matrix $R$, whereas the decohering term causes the decay of off-diagonal elements (coherences).
To model a decohering environment we set all $R_{ij}=0$, thereby suppressing thermally induced transitions between energy eigenstates.
This leads immediately to Eq.~\ref{DME}.

Earlier, we had motivated our definition of work in the presence of a decohering environment, $W = \bar{\epsilon}_m - \epsilon_n$, heuristically.
With Eq.~\ref{DME} this argument can be strengthened using a simple microscopic model, as we describe in the Appendix 5.2.

Note that evolution under Eq.~\ref{DME} preserves the identity, $\mathcal{L} \hat{I} = 0$, hence this evolution is {\it unital}, and Eq.~\ref{EQ:JE} follows as an immediate consequence of a general result derived by Rastegin~\cite{Rastegin2013}.
To keep our presentation self-contained, we now derive Eq.~\ref{EQ:JE} assuming only a linear master equation that preserves the identity.

Let $\Lambda_\tau: \hat\rho_0 \rightarrow \hat\rho_\tau$ denote the quantum evolution that maps an initial density matrix to a final density matrix, under the dynamics of Eq.~\ref{DME}.
After initial equilibration, an energy measurement at time $t=0$ yields an energy eigenvalue $\epsilon_n$ with probability $p_n = Z_0^{-1}e^{-\beta\epsilon_n}$, and ``collapses'' the system into a pure state $\hat\rho_0 = \vert n \rangle \langle n \vert$.  This state then evolves under Eq.~\ref{DME} to $\hat\rho_\tau = \Lambda_\tau(\hat\rho_0)$ and a final energy measurement at $t=\tau$ yields a value $\bar\epsilon_m$ with probability $p_{\bar m\vert n} = \langle \bar{m} \vert \hat\rho_\tau \vert\bar{m}\rangle$.
Summing over all possible measurement outcomes, and using the linearity and identity preservation of $\Lambda_\tau$, we have~\cite{Rastegin2013}
\begin{eqnarray*}
\langle e^{-\beta W} \rangle
&= \sum_{nm} p_n \, p_{\bar m\vert n} \, e^{-\beta (\bar{\epsilon}_m - \epsilon_n)} \\
&=\sum_{nm} \frac{e^{-\beta \epsilon_n}}{Z_0} \langle \bar{m}|\Lambda_\tau(|n \rangle \langle n|)|\bar{m}\rangle e^{-\beta(\bar{\epsilon}_m - \epsilon_n)} \\
&= \frac{1}{Z_0}\sum_{m} e^{-\beta \bar{\epsilon}_m} \langle \bar{m}|\Lambda_\tau(\hat{I})|\bar{m}\rangle =\frac{Z_\tau}{Z_0} = e^{-\beta \Delta F} \quad .
\end{eqnarray*}

\begin{figure}[ht]
\centering
\includegraphics[width=0.75\textwidth]{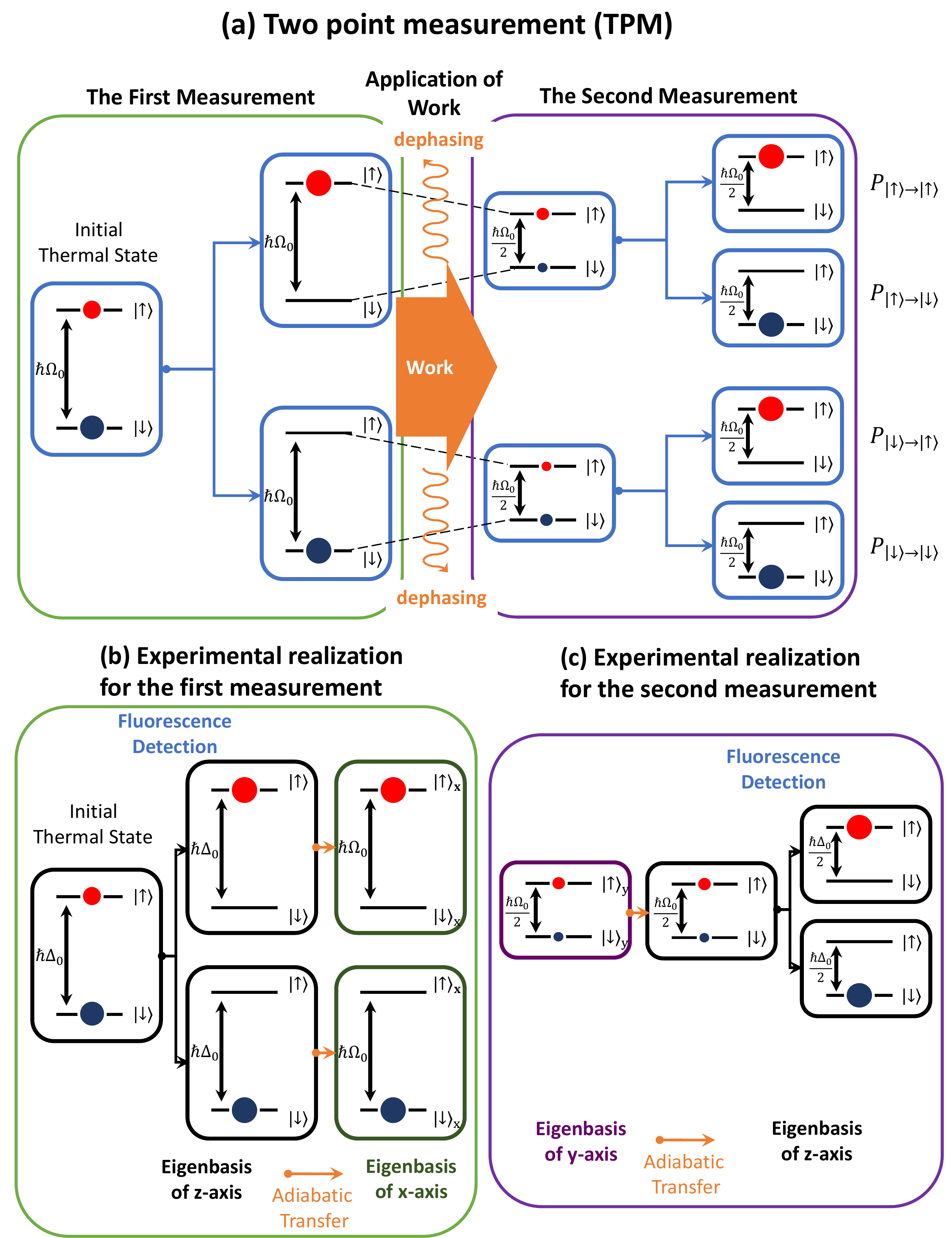}
\caption{(a) and (b)(c) respectively show conceptual and actual experimental schematics of the TMP protocol in our setup. (b) indicates that in the true experiment thermal state preparation and initial energy measurement occur in the $\hat{\sigma}_{\rm z}$ eigenbasis before being transfered to the basis of $\hat{\sigma}_{\rm x}$ with the aid of an adiabatic shortcut. (c) indicates how the system is again rotated--this time from the $\hat{\sigma}_y$ to $\hat{\sigma}_z$ basis--proceeding the second fluorescence measurement. Note that the level splitting in the $\hat{\sigma}_{\rm z}$ basis is set by $\Delta_{0}$ which is the frequency difference between the laser beat-note and $\omega_{0}$. }\label{fig:Scheme}
\end{figure}

\section{Experimental Verification}
To test Eq. (\ref{EQ:JE}) experimentally, we employ a two state system engineered from a \Yb ion's orbital degrees of freedom, using the energy levels $\ket{F=0,m_F=0}\equiv\ket{\downarrow}$ and $\ket{F=1,m_F=-1}\equiv\ket{\uparrow}$ belonging to the ground-state manifold of $^2$S$_{1/2}$ \cite{Xiang13}.  By applying microwave pulses resonant to our states' energy difference $\omega_{\rm 0}\equiv\omega_{\rm HF} - \omega_{\rm Z}$, where $\omega_{\rm HF}=\left(2\pi\right)12.642821{\rm GHz}$ and $\omega_{\rm Z}=\left(2\pi\right)13.586 {\rm MHz}$, the system can be driven according to the Hamiltonian
\begin{equation}
\hat{H}(t) =  \frac{\hbar\Omega(t)}{2} \left[ \hat{\sigma}_{\rm x} \cos \phi(t)  +\hat{\sigma}_{\rm y} \sin \phi (t)\right].
\label{eq:Ham}
\end{equation}
Here $\hat{\sigma}_{x,y}$ are the standard Pauli matrices in the $\{ \ket{\uparrow} , \ket{\downarrow} \}$ basis
while $\Omega$ and $\phi$ are parameters controlled through the amplitude and phase of the microwave pulses.  In our experiment, we use the driving protocols
\begin{equation}
\Omega(t) = \Omega_{0} \left( 1- \frac{t}{2 \tau}\right) \,\,\,;\,\,\,\phi(t) = \frac{\pi t}{2 \tau}
\label{eq:Proto}
\end{equation}
where $\tau$ is the duration of the process.  Together equations (\ref{eq:Ham}) and (\ref{eq:Proto}) represent the Hamiltonian portion of our system's dynamics.  The decohering term of Eq. (\ref{DME}) is realized by the addition of noise in the microwave pulse sequence.  In our setup this adds a stochastic term $\Omega_0 \xi(t)$ to the protocol $\Omega(t)$ where $\xi(t)$ is gaussian white noise characterized by zero mean $\avg{\xi(t)}=0$ and variance $\avg{\Delta\xi(t)\xi(t+\tau)}= \alpha^{2} \delta(\tau)$.  Averaging over all realizations of the noise $\xi(t)$ produces an equation of motion identical to Eq. (\ref{DME}) with $\gamma_{ij} = \gamma = \frac{1}{2} \alpha^{2} \Omega_{0}^2$ \cite{Loreti94,Biercuk14} (see also Appendix 5.4).

\begin{figure}[ht]
\centering
\includegraphics[width=1.0\textwidth]{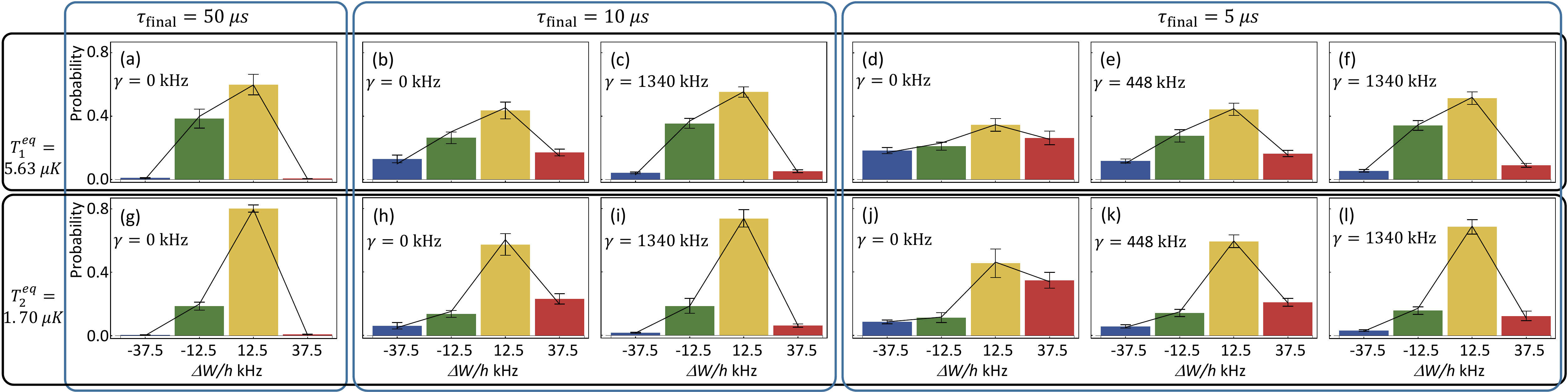}
\caption{The work distributions (a)-(f) correspond to an initial temperature of $T_{1}=5.63$ $\mu$K while (g)-(l) have $T_{2}=1.70$ $\mu$K.  The driving times $\tau=50 \mu s$, $\tau=10 \mu s$, and $\tau = 5 \mu s$ represent near adiabatic (a)(g), moderate(a)(c)(h)(i), and fast (d)(e)(f)(j)(k)(i) driving regimes. The dephasing rate $\gamma$ took values of $0$, $448$, and $1340$ kHz for the cases of no (a)(b)(d)(g)(h)(j), intermediate (e)(k), large (c)(f)(i)(l) dephasing respectively.}\label{fig:WorkDistribution}
\end{figure}

Given this setup, the procedure for measuring the work applied during a single experimental trial involves four steps: $(i)$ thermal state preparation, $(ii)$ initial energy measurement, $(iii)$ application of the driving protocol, and $(iv)$ final energy measurement, as shown in Fig. \ref{fig:Scheme}(a).

Our Hamiltonian has the form $\hat H(t) = {\bf B}(t) \cdot \hat{\boldsymbol\sigma}$, where the field ${\bf B}(t)$ undergoes rotation by $90^\circ$ in the $xy$-plane (see Eq.~\ref{eq:Ham}).
For technical reasons the initial thermalization and both measurements are performed in the $\hat{\sigma}_z$ basis. Therefore after the initial thermalization and measurement we rotate the system from the $z$-axis into the $xy$-plane, then we implement the driving as per Eq.~\ref{eq:Ham}, and finally we rotate the system back to the $z$-axis to perform the final measurement. These rotations do not affect the work distribution. The rotations are achieved with \textit{adiabatic shortcuts} \cite{Rice03,Berry09,An16}, which produce transformations equivalent to adiabatically switching the system's Hamiltonian, but in a finite time (See Appendix 5.7). Fig. \ref{fig:Scheme} (b)(c) show detailed schematics of the measurement protocols, including these shortcuts.

\textit{(i) Thermal state preparation} - We create the initial thermal state using the following procedure.  First we prepare the pure state $\ket{\psi} = c_{\uparrow}\ket{\uparrow} + c_{\downarrow}\ket{\downarrow}$ using a standard optical pumping sequence followed by the application of resonant microwaves over a proper duration. After waiting more than 10 times the coherence time (see Appendix 5.5), the state becomes a mixed-state described by the density operator $\hat{\rho}_{\rm ini}=|c_{\uparrow}|^2\ket{\uparrow}\bra{\uparrow} + |c_{\downarrow}|^2\ket{\downarrow}\bra{\downarrow}$, which is identical to thermal equilibrium state $\exp(-\hat{H}(0)/k_{\rm B}T)$ with an effective temperature
\begin{equation}
T = \frac{\hbar \Omega_{\rm 0}}{k_{\rm B} \ln(|c_{\downarrow}|^2/|c_{\uparrow}|^2)}.
\end{equation}
For our experiment, $\Omega_{\rm 0}=2\pi\times50$ kHz while $|c_{\downarrow}|^2$ took values of $0.605 \pm 0.041$ and $0.804 \pm 0.034$, corresponding to effective initial state temperatures of $T_1 = 5.63$ $\mu \rm K$ and $T_2 = 1.70$ $\mu \rm K$, respectively.

\textit{(ii) Initial energy measurement} - Following initial state preparation, the energy of the system is measured using a standard state-sensitive fluorescence detection sequence. In this procedure, fluorescence or the absence of fluorescence during the detection sequence indicate a measurement of the $\ket{\uparrow}$ or $\ket{\downarrow}$ state respectively.  When the ground state $\ket{\downarrow}$ (dark state) is measured, we continue to the next step of the experiment. If the excited state $\ket{\uparrow}$ (bright state) is detected, we re-prepare the $\ket{\uparrow}$ state before continuing (see Appendix 5.6).  As noted above, the actual measurements are performed with respect to the Hamiltonian $\hbar \Omega_0 \hat{\sigma}_z / 2$ which is then switched to $\hbar \Omega_0 \hat{\sigma}_x / 2$ using an adiabatic shortcut (see Appendix 5.7).

\textit{(iii) Application of driving with dephasing} - At this point noisy microwave pulses are applied to the system resulting in evolution according to the Hamiltonian (\ref{eq:Ham}) with the protocols (\ref{eq:Proto}) and decoherence.  For our trials, $\tau$ took values $50 \mu s$, $10 \mu s$, and $5 \mu s$ representing near adiabatic, intermediate, and fast driving speeds.  The decoherence rate $\gamma$ in Eq. (\ref{DME}) was set to $0$, $448$, or $1340$ kHz which correspond to the cases of no, intermediate, or large dephasing strength respectively.

\textit{(iv) The final energy measurement} - Prior to the final energy measurement, another adiabatic shortcut is used to switch the system's Hamiltonian--this time from $\hbar \Omega_0 \hat{\sigma}_y / 4$ to  $\hbar \Omega_0 \hat{\sigma}_z / 4$.  Following this transfer, the energy of the system is once again measured using a state-sensitive fluorescence detection sequence.  By calculating the difference between the initial and final energy measurements, a work value for the experimental trial is obtained.

Figure \ref{fig:WorkDistribution} shows the work distributions resulting from experiments conducted with twelve different combinations of effective temperature $T$, driving time $\tau$, and decoherence rate $\gamma$.  From the data, it is clear that decoherence non-trivially affects the work distribution for a given process -- for instance compare (d) - (f) in Fig. \ref{fig:WorkDistribution}.  A more careful inspection reveals that the qualitative behavior of the work distribution is governed by a competition between driving speed and decoherence.  For near-adiabatic driving, the work distribution is peaked at values $W = \bar{\epsilon}_i - \epsilon_i$ corresponding to the measurement of two energies with the same quantum number.  Increasing driving speed (decreasing $\tau$) tends to induce transitions among energy states with different quantum numbers, thereby broadening the work distribution.  This effect is exemplified in Fig. \ref{fig:WorkDistribution} by distributions (a), (b), and (d).  In contrast, decoherence in the eigenbasis of $\hat H(t)$ suppresses these transitions bringing the work distribution closer to its adiabatic form.  This can be seen by comparing the near adiabatic distribution (a) with the fast driving cases (d),(e), and (f) which have varying degrees of decoherence.  Interpreting this decoherence as environmental measurement of the system's energy, one can see that the system is forced to follow the adiabatic trajectory due to wave function collapse.  When the collapse rate $\gamma$ becomes large, the system becomes trapped in an eigenstate of the instantaneous Hamiltonian -- a scenario analogous to the quantum Zeno effect.

\begin{figure}[ht]
\centering
\includegraphics[width=\textwidth]{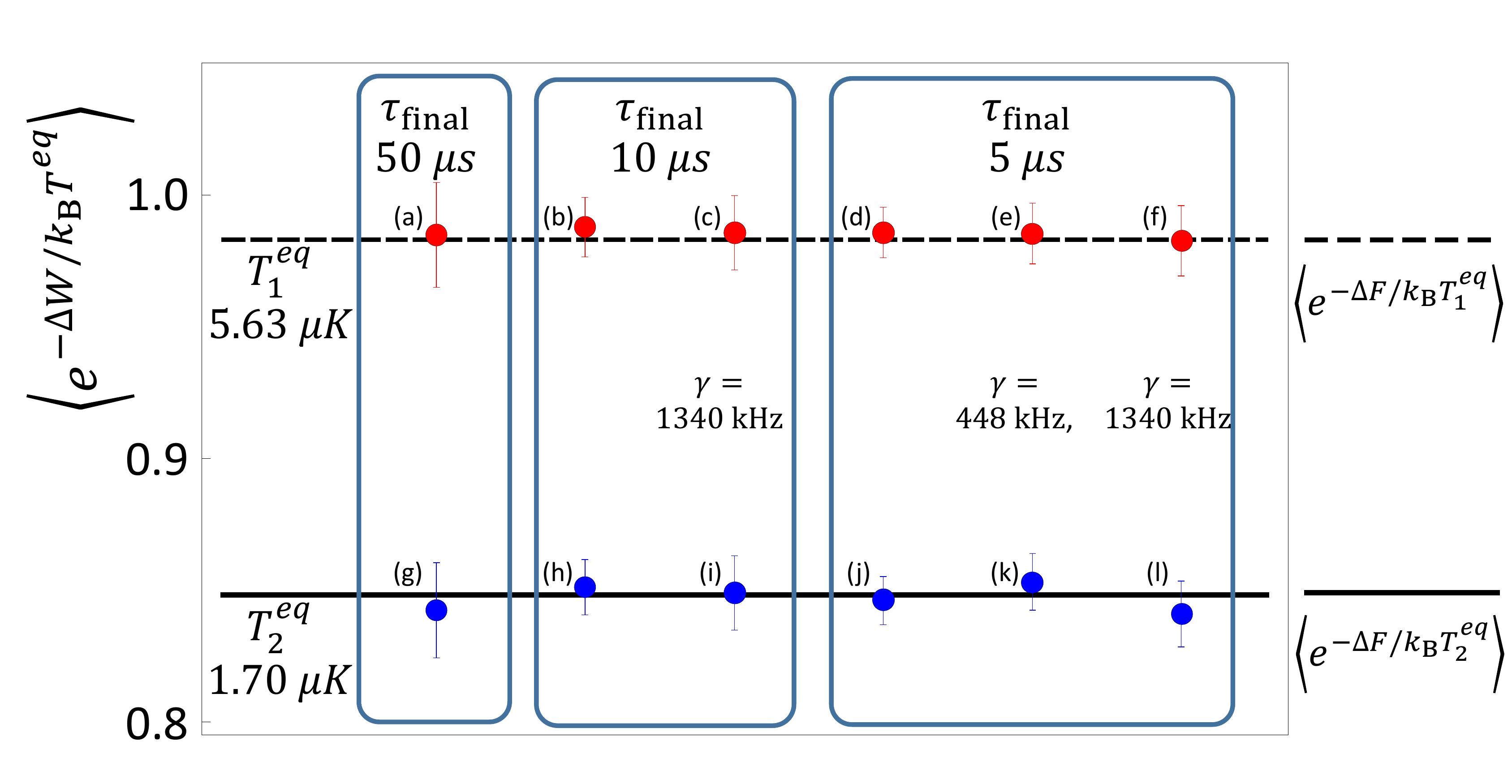}
\caption{Comparison of the exponential average of work for distributions (a)-(l) in Fig. \ref{fig:WorkDistribution} to the exponential of the free energy difference calculated from the initial and final energy levels of $\hat{H}(t)$.}\label{fig:QWR}
\end{figure}

With these distributions, the work relation can be tested for each choice of the experimental parameters $T$, $\tau$, and $\gamma$ by direct comparison of the left and right hand sides of Eq. (\ref{EQ:JE}).  Note that the quantity $\langle e^{-\beta W} \rangle$ is calculated using the work distribution while  $e^{-\beta \Delta F}$ follows straightforwardly from knowledge of the energy levels of $\hat{H}(0)$ and $\hat{H}(\tau)$.  The results of these calculations, shown in Fig. \ref{fig:QWR}, agree to within the error of the experiment and hence validate the work relation.

\section{Discussion and Conclusions}
While our results focus on systems that solely experience decoherence, significant theoretical progress has been made in understanding the quantum work relation for situations where dissipation is also important.  We  outline some of these advances as they give context for this manuscript and provide direction for future experimental tests of quantum fluctuation theorems.

Perhaps the most conceptually appealing framework that addresses general thermal environments is based on considering the system and environment jointly as a closed composite system \cite{Talkner2011, Deffner2011}.  Here the TPM scheme can be employed as the work is simply the change in energy of the joint system.  (In the weak coupling limit, work can also be defined as $\Delta U - Q$ where the energy change $\Delta U$ and the heat $Q$ are obtained by applying the TPM protocol separately to the system and environment.)  Despite defining a work distribution that satisfies Eq. (\ref{EQ:JE}), this approach suffers from the need to measure bath degrees of freedom, which is difficult to realize in practice.

Other studies of the work relation overcome this issue by defining work at the system level without referencing an environment.  In this vein there are several equivalent formalisms for treating quantum detailed balance master equations \cite{Liu2012,Chetrite2012,Liu2014a,Liu2014b,Liu2016,Esposito2009,Silaev2014} of which we focus on the quantum jump trajectory method~\cite{Horowitz2012,Horowitz2013,Breuer2013,Pekola2013,Liu2014a,Liu2014b,Jordan2015}.   Originally developed in the field of quantum optics \cite{Dalibard1992}, this approach treats a system's density operator as an average over pure states evolving according to stochastic trajectories.  The construction of these trajectories is called an \textit{unraveling} and is generally not unique.  When this unraveling is chosen properly \cite{Elouard2015}, it can be shown that a consistent trajectory-based thermodynamics can be defined in a manner similar to classical stochastic thermodynamics and that the work relation remains valid \cite{Horowitz2012,Horowitz2013,Breuer2013,Pekola2013,Liu2014a,Liu2014b,Jordan2015}.  When applied to the decohering master equation (\ref{DME}), the quantum trajectory approach agrees with the theoretical development section of this paper.

Although we have used Eq.~\ref{DME} to model a system in weak contact with an environment, the same equation (but setting $\gamma_{ij}=\gamma$ for all $i\ne j$) describes a system that undergoes unitary evolution under $\hat H(t)$, interrupted by projective energy measurements performed at random times $t_1, t_2, \cdots$ that are Poisson-distributed at a rate $\gamma$.
In this alternative scenario the projective measurements produce stochastic energy changes that are described as {\it quantum heat} ($Q_q$) in the framework of Elouard {\it et al}~\cite{Elouard2017}.
In the present paper we have interpreted the quantity $\bar{\epsilon}_m - \epsilon_n$ as the work ($W$) performed on the system;
in Ref.~\cite{Elouard2017} the same quantity is interpreted as the sum of work and quantum heat ($W+Q_q$).
It would be interesting to study an experimental situation in which the two contributions, work and quantum heat (as defined in Ref.~\cite{Elouard2017}), could be determined separately.

Various approaches might be taken in future experimental tests of quantum fluctuation theorems.  For instance, rather than producing decoherence through the addition of noise, a true decohering bath could be engineered using an interaction commuting with the bare Hamiltonians of the system and environment.   Additionally, the quantum work relation could be tested for a general thermal environment using the TPM protocol and a continuous environmental measurement technique \cite{Wiseman1993,Wiseman2010,Breuer2002, Pigeon2016} such as single photon detection in a cavity QED experiment.  Alternatively using only the TPM protocol on a dissipative system, one could test the energy change fluctuation theorem which is a modified version of equation (\ref{EQ:JE}) devised by Pekola and co-workers \cite{Pekola2015}.

In summary, we have studied the quantum work relation for a system in contact with a decohering bath.  We obtained Eq.~(\ref{EQ:JE}) within a simple, phenomenological model that complements the more general approaches of unital quantum channels and quantum trajectories. Using a system constructed from trapped ions, we conducted an experiment that verified the work relation for a decohering process and represents the first test of Eq. (\ref{EQ:JE}) beyond the regime of closed quantum systems.  These results demonstrate the applicability of fluctuation theorems to open quantum systems, at least for the special case of a decohering heat bath, and may spur additional tests of the work relation for systems with dissipation.

\section{APPENDIX}
\subsection{Detailed Balance Master Equation}
Consider a quantum detailed balance master equation with a Hamiltonian $\hat{H} = \sum \epsilon_i | i \rangle \langle i |$ and an equilibrium state $\hat{\rho}^{eq}$ satisfying the standard thermal relation
\begin{equation}
\hat{\rho}^{eq} = \frac{e^{-\beta \hat{H}}}{Tr[e^{-\beta \hat{H}}]}.
\end{equation}
Additionally assume that the gaps $\epsilon_i-\epsilon_j$ in the spectrum of $\hat{H}$ are non-degenerate.
Under these conditions, Alicki showed \cite{Alicki1976} that the master equation may be written in the form
\begin{equation}
\frac{d \hat{\rho}}{dt} = -\frac{i}{\hbar} [\hat{H},\hat{\rho}] + \sum_{i,j=1}^N D_{ij} \Big\{[\hat{X}_{ij},\hat {\rho}\hat{X}_{ij}^{\dagger}] + [\hat{X}_{ij}\hat{\rho},\hat{X}_{ij}^{\dagger}]\Big\} \label{dbme}
\end{equation}
where $N$ is the dimension of the system's Hilbert space and the real numbers $D_{ij}$ and operators $\hat{X}_{ij}$ satisfy the conditions
\begin{equation}
D_{ij} e^{-\beta\epsilon_j} = D_{ji}e^{-\beta\epsilon_i} \,\,\, ; \,\,\, D_{ij} \geq 0 \label{dbcond1}
\end{equation}
\begin{equation}
[\hat{H},\hat{X_{ij}}] = (\epsilon_i - \epsilon_j) \hat{X}_{ij} \label{dbcond2}
\end{equation}
\begin{equation}
Tr[\hat{X}_{ij}^\dagger \hat{X}_{kl} ] = \delta_{ik} \delta_{jl} \label{dbcond3}
\end{equation}
\begin{equation}
\hat{X}_{ij} = \hat{X}_{ji}^\dagger \label{dbcond4}.
\end{equation}
In what follows, we will use the non-degenerate gaps of $\hat{H}$ along with conditions (\ref{dbcond1})-(\ref{dbcond4}) to gain insight into the constants $D_{ij}$ and operators $\hat{X}_{ij}$. This in turn will allow for equation (\ref{dbme}) to be written in a form where the processes of relaxation and decoherence are manifest.

\textit{Constants $D_{ij}$} -- The constants $D_{ij}$ can largely be interpreted within the framework of a classical continuous time Markov process \cite{VKampen2007}.  Assuming discrete states indexed by $i$, such processes describe the evolution of a probability distribution $p_i$ according to
\begin{equation}
\frac{dp_i}{dt}= \sum_j r_{ij}p_j
\end{equation}
where $r_{ij}$ is a transition rate matrix with the properties
\begin{equation} \label{trm}
r_{ij} \left\{\begin{array}{lll}
\geq 0 ; &\,\, (i\neq j) \\ \\
= -\displaystyle{\sum_{k \neq i}} r_{ki} &\,\, (i = j).
\end{array}\right.
\end{equation}
Furthermore the matrix $r_{ij}$ is said to satisfy detailed balance with respect to an equilibrium probability distribution $p^{eq}_i$ when
\begin{equation}
r_{ij} p^{eq}_j - r_{ji}p^{eq}_i = 0. \label{dbclass}
\end{equation}

Given these definitions, one immediately recognizes from (\ref{dbcond1}) that the off diagonal elements of $D_{ij}$ coincide with the elements of a transition rate matrix satisfying the detailed balance condition (\ref{dbclass}) with $p^{eq}_i \propto \exp (-\beta \epsilon_i)$. 
In what follows, we will find that the energy populations $\rho_{ii} = \langle i | \hat{\rho} | i \rangle$ relax thermally according to
\begin{eqnarray}
\frac{d\rho_{ii}}{dt} &=& \sum_{j \neq i} (2 D_{ij}) \rho_{jj} + (-2 \sum_{j \neq i}D_{ji})\rho_{ii} \\ \nonumber \\
&=& \sum_{j \neq i} r_{ij} \rho_{jj} + r_{ii} \rho_{ii}. \nonumber
\end{eqnarray}
Hence for $i \neq j$ we will interpret $D_{ij}$ as half the thermally induced transition rate from energy state $j$ to state $i$.  Note that $r_{ii}$ is defined according to (\ref{trm}) and $D_{ii} \neq r_{ii}/2$.  Condition (\ref{dbcond1}) only constrains the constants $D_{ii}$ to be positive.  These numbers will later be interpreted in terms of decoherence rates.  Anticipating these connections, the elements of $D_{ij}$ will be redefined according to
\begin{equation} \label{dij}
D_{ij} = \left\{\begin{array}{lll}
 r_{ij}/ 2 &\,\, (i\neq j) \\ \\
 d_i &\,\, (i = j).
\end{array}\right.
\end{equation}

\textit{Operators $\hat{X}_{ij}$} -- Before finding the explicit form of the operators $\hat{X}_{ij}$, it is instructive to recast conditions (\ref{dbcond2}) and (\ref{dbcond3}) in the language of linear algebra.  Specifically note that (\ref{dbcond2}) dictates that $\hat{X}_{ij}$ is an eigen-operator of the super-operator $[\hat{H}, \cdot]$ with eigenvalue $\epsilon_i - \epsilon_j$ while (\ref{dbcond3}) asserts that the operators $\hat{X}_{ij}$ form an orthonormal set with respect to the matrix inner product $\langle \hat{A}, \hat{B} \rangle = Tr[\hat{A}^\dagger \hat{B}]$.

First consider the operators $\hat{X}_{ij}$ for which $i \neq j$.  In this case, each eigenvalue $\epsilon_i-\epsilon_j$ of equation (\ref{dbcond2}) is non-degenerate (due to the gap structure of $\hat{H}$) and hence the corresponding eigen-operator $\hat{X}_{ij}$ is confined to a one dimensional eigenspace.  By inspection this eigenspace is determined to be $\{ \alpha | i \rangle \langle j | : \alpha \in \mathbb{C} \}$.  The normalization condition (\ref{dbcond3}) further gives the constraint that $|\alpha|^2 = 1$.  Without loss of generality, it is now possible to set
\begin{equation}
\hat{X}_{ij} = | i \rangle \langle j | \,\,\,\,\,\, (i \neq j) \label{Xij1}
\end{equation}
due to the fact that the master equation (\ref{dbme}) is independent of the phase of $\alpha$ since  $\hat{X}_{ij}$ and $\hat{X}^\dagger_{ij}$ appear in conjugate pairs.

For the case where $i=j$, the eigenvalue in equation (\ref{dbcond2}) vanishes and corresponds to the $N$ dimensional eigenspace $\{\sum_k O_{ik} | k \rangle \langle k | : O_{ik} \in \mathbb{C} \} $.  Application of conditions (\ref{dbcond3}) and (\ref{dbcond4}) gives
\begin{equation}
O_{ik} \in \mathbb{R} \,\,\,;\,\,\, \sum_{k} O_{ik} O_{jk} = \delta_{ij}
\end{equation}
which is exactly the condition that the matrix $O_{ik}$ belong to the set of real orthogonal matrices $O(N)$.  In conclusion
\begin{equation}
\hat{X}_{ii} = \sum_k O_{ik} | k \rangle \langle k | \,\,\,;\,\,\, O_{ik} \in O(N). \label{Xij2}
\end{equation}

The form of the detailed balance master equation in the main body of this manuscript can now be deduced.  Following substitution of (\ref{dij}), (\ref{Xij1}), and (\ref{Xij2}) into the master equation (\ref{dbme}) and some manipulation,  the result is given by
\begin{equation} \label{QDBMEA}
\frac{d \hat{\rho}}{dt} = -\frac{i}{\hbar}[\hat{H},\hat{\rho}] + \sum_{ij} J_{ij} |i\rangle \langle i | + \sum_{i\ne j} \Gamma_{ij}  |i\rangle \langle j |  
\end{equation}
\begin{equation*} 
J_{ij} \equiv r_{ij} \rho_{jj} - r_{ji} \rho_{ii} 
\end{equation*}
\begin{equation*} 
\Gamma_{ij} \equiv [(r_{ii} + r_{jj})/2 - \gamma_{ij}] \rho_{ij} \leq 0 
\end{equation*}
\begin{equation*} 
\gamma_{ij} \equiv \sum_k d_k (O_{ki} - O_{kj})^2 \geq 0. 
\end{equation*}
As stated earlier, the virtue of writing the master equation in the above form is that the processes of relaxation and decoherence are clearly displayed -- they are the second and third terms on the right hand side of (\ref{QDBMEA}) respectively.  The relaxation is seen to shuffle the diagonal elements of the density operator according to a Markov process while the decoherence term causes exponential decay of off-diagonal elements.

\subsection{The decohering master equation from a Hamiltonian model}

In our main theoretical development, we argued that it is plausible no heating occurs during a decohering process and hence it is reasonable to determine work values using the two-point measurement protocol.  Here we strengthen this argument by presenting a specific microscopic model where our intuition can be verified according to the definitions of heat and work presented by Campisi et al \cite{Talkner2011}.

Specifically, we consider a simple repeated interaction model where the bath is represented by a stream of identical auxiliary systems which we will refer to as \textit{units}.  Each unit begins in a thermal state $\hat{\omega}$ and interacts with the system of interest for a time $\delta t$.  Over every interaction interval, the total Hamiltonian (system plus units) is fixed but the system's Hamiltonian and the interaction may change suddenly between intervals.  We will denote the total Hamiltonian during the nth interval by
\begin{equation}
\hat{H}_n = \hat{H}_n^{(S)} \otimes \hat{I}^{(U)} + \hat{I}^{(S)} \otimes \hat{H}^{(U)} + \lambda \hat{V}_n
\end{equation}
where $\hat{H}_n^{(S)}$ is the system's Hamiltonian, $\hat{H}^{(U)}$ is the Hamiltonian of the non-interacting units which each have individual Hamiltonians $\hat{h}^{(U)}$, $\lambda$ is the interaction strength, and $\hat{V}_n$ is an interaction that acts only on the system and nth unit.  Furthermore to assure the process only produces dephasing in the system of interest, we assume that the interaction is of the form
\begin{equation}
\hat{V}_n = \hat{A}_n \otimes \hat{B}
\end{equation}
where $\hat{A}_n$ acts on the system and commutes with $\hat{H}_n^{(S)}$ while $\hat{B}$ acts on the nth unit and commutes with $\hat{h}^{(U)}$.  In the following, we outline two important properties of this model: 1) the existence of a regime where the system's dynamics are described by a decohering master equation and 2) the absence of heat transfer between the system and units.

In order to show 1), we take
\begin{eqnarray}
\hat{H}_n^{(S)} &=& \hat{H}^{(S)}(n \delta t) \\
\hat{A}_n &=& \hat{A}(n \delta t)
\end{eqnarray}
where $\hat{H}^{(S)}(t)$ and $\hat{A}(t)$ are operators that vary continuously with time and make the standard assumption \cite{Breuer2002} that $Tr[\hat{\omega} \hat{B}]=0$.  Taking the limit $\delta t \to 0$ while simultaneously letting the interaction strength grow according to $\lambda = k \delta t^{-1/2}$ where $k$ is a positive real constant, it can be shown \cite{Strasberg17} that
\begin{eqnarray}
\frac{d \rho^{(S)}}{dt} = &-&\frac{i}{\hbar}[\hat{H}^{(S)}(t), \rho^{(S)}]  \\ &-& C\left[\hat{A}(t)\rho^{(S)}\hat{A}(t) - \frac{1}{2} \{\hat{A}^2(t),\rho^{(S)} \}\right] \nonumber \label{RIME}
\end{eqnarray}
\begin{equation*}
C = \frac{2 k \textrm{Tr}[\hat{B}^2 \hat{\omega}]}{\hbar^2} 
\end{equation*}
Since $\hat{H}(t)$ and $\hat{A}(t)$ commute at all times, they share a common eigenbasis $\{ | i (t) \rangle \}$.  Rewriting the dissipator (second term on the RHS of (\ref{RIME})) in  in this basis, the master equation becomes
\begin{eqnarray}
\frac{d \rho^{(S)}}{dt} = - \frac{i}{\hbar}[\hat{H}^{(S)}(t), \rho^{(S)}]  \\ - \sum_{i \neq j} \gamma_{ij} |i(t)\rangle \langle i(t)| \rho^{(S)} |j(t)\rangle \langle j(t)| \nonumber
\end{eqnarray}
\begin{equation*}
\gamma_{ij} = \frac{\textrm{Tr}[\hat{B}^2 \hat{\omega}]}{\hbar^2}(a_i -a_j)^2 
\end{equation*}
where $a_i$ are the eigenvalues of $\hat{A}$.

We now show property 2) holds according to the definitions of heat and work proposed in \cite{Talkner2011}.  In this setup, work is determined (for initially thermal states)  by applying the two point measurement protocol to the joint system and environment.  Assuming that the system is decoupled from the units at the beginning and end of the process, the work performed during a single realization is given by $W = \epsilon^{(S)}_m + \epsilon^{(U)}_k - \epsilon^{(S)}_n - \epsilon^{(U)}_l$ where $\epsilon^{(S)}_m + \epsilon^{(U)}_k$ and $\epsilon^{(S)}_n + \epsilon^{(U)}_l$ respectively are the initial and final energy measurements.  Since the total Hamiltonian of the system and units commutes with $\hat{H}^{(U)}$ at all times, it follows that $\epsilon^{(U)}_k=\epsilon^{(U)}_l$ which implies that the work is fully determined by local measurements on the system of interest as claimed in the main text of this manuscript.   

\subsection{Quantum work with a decohering environment}

In the main text, we argued that the two-point measurement protocol provides a proper definition of the quantum work performed on a system in contact with a decohering environment: $W = \bar{\epsilon}_m - \epsilon_n$.
Our argument was essentially heuristic: {\it by assumption} there is no exchange of energy (no dissipation) therefore all changes in the system's energy are attributed to work.
Here we provide a more quantitative argument, using the master equation introduced in the main text:
\begin{equation}
\label{DMEA}
\frac{d\hat\rho}{dt} = -\frac{i}{\hbar}[\hat{H}(t), \hat\rho ] - \sum_{i\neq j} \gamma_{ij} \rho_{ij}  |i\rangle \langle j |  \quad.
\end{equation}

First, let us momentarily imagine that the system begins in a non-diagonal density matrix (in the energy basis) and then evolves under Eq.~\ref{DMEA}, with the Hamiltonian {\it held fixed}. In this situation the diagonal elements of $\rho_{ii}(t)$, and hence the energy distribution, remain unaffected.  Therefore there is no exchange of energy (heat) between the system and the bath, only a decay of coherences as the off-diagonal elements decay at rates $\gamma_{ij}$.

Now imagine the following scenario: the system first evolves for a short interval of time $\delta t$ under {\it unitary} dynamics, with time-dependent $\hat H(t)$; then the Hamiltonian is held fixed and the system evolves with decoherence, as in the previous paragraph, for a time interval $\delta t$; then again under unitary evolution with time-dependent $\hat H(t)$; followed by decoherence with fixed $\hat H$; and so forth.  In other words, the system alternates between brief intervals of either time-dependent $\hat H(t)$ or decoherence, but not both simultaneously.  Representing this evolution by a density matrix $\hat \rho(t)$ in the instantaneous eigenbasis of $\hat H(t)$, we note that during the unitary intervals with time-dependent $\hat H(t)$, the diagonal elements of $\hat \rho(t)$ change with time, but during the intervals of decoherence at fixed $\hat H$ they remain constant.  In other words, the energy distribution changes only during the unitary intervals, reflecting the performance of work, and not during the decohering intervals, reflecting the absence of heat. Of course, the presence of the decohering intervals affects the final energy distribution (as demonstrated in Fig. \ref{fig:Scheme}), simply because the off-diagonal elements of $\hat \rho$ at the start of a given unitary interval are affected by the amount of decoherence during the preceding decohering interval; but this does not imply that energy (heat) is being exchanged with the environment.

We can view our experimental setup as a limiting case of the scenario described in the previous paragraph, in which $\delta t\rightarrow 0$.

Alternatively, we can imagine that $\hat H(t)$ undergoes small, sudden changes at equally spaced times $t_n=n\, \delta t$, but remains constant in between these steps. At these discrete times $\{t_n\}$, the state of the system does not instantaneously change, but the elements of $\hat \rho(t)$ change abruptly, as the eigenbasis of $\hat H$ undergoes small rotations. These changes are associated with work, just as in the case of closed quantum systems.  In between these steps, as the system evolves under Eq. 4, with fixed $\hat H$, there is decoherence but no heat.  As we let $\delta t\rightarrow 0$, the discrete changes in $\hat H$ become infinitesimal, and we recover the situation described in our paper.

\subsection{Stochastic noise and Decoherence rate}
In our experiment, decoherence is induced by the introduction of noise. The system is driven by the total Hamiltonian
\begin{equation}
\hat{H}(t) = \frac{\hbar [\Omega(t) + \Omega_0 \xi(t)]}{2}\hat{\sigma}_{\vec{n}}(t)
\end{equation}
where $\hat{\sigma}_{\vec{n}}(t) = \hat{\sigma}_x \cos{\phi(t)} + \hat{\sigma}_y \sin{\phi(t)}$ and $\xi(t)$ is Gaussian white noise characterized by $\langle \xi(t) \rangle = 0$ and $\langle \xi(t)\xi(t+\tau) \rangle = \alpha^2 \delta(\tau)$. $\hat{H}(t)$ can be decomposed into a control part $\hat{H}_c(t) = \hbar\Omega (t) \hat{\sigma}_{\vec{n}}(t)/2$ and stochastic part $\hat{H}_s(t) = \hbar\Omega_0 \xi(t) \hat{\sigma}_{\vec{n}}(t)/2$.

Taking the ensemble average over all noise realizations, the evolution of the system is described by the Lindblad master equation\cite{Loreti94,Band15,VKampen2007}
\begin{equation}
\frac{d\hat{\rho}}{dt} = -\frac{i}{\hbar}[\hat{H}_c(t), \hat{\rho}] - \gamma (\rho_{\downarrow\uparrow} \ket{\downarrow} \bra{\uparrow} + \rho_{\uparrow\downarrow}\ket{\uparrow}\bra{\downarrow})
\end{equation}
where $\ket{\uparrow}, \ket{\downarrow}$ are the instantaneous eigenvectors of $\hat{H}_c(t)$ and $\gamma$ is the decoherence rate which satisfies
\begin{equation}
\gamma = \frac{(\alpha\Omega_0)^2}{2}. \label{decayrate}
\end{equation}

In practice we applied discrete noise with a sampling rate of $R_{\rm s}$ instead of ideal continuous-Gaussian white noise. When  $R_{\rm s}^{-1}/2$ is much less than the duration of the operation, the digital noise can be approximated as Gaussian white noise, with auto-correlation function $\langle \xi(t)\xi(t+\tau) \rangle = \sigma^2 R_{\rm s}^{-1} \delta(\tau)$. Hence Eq.(\ref{decayrate}) should be revised as
\begin{equation}
\gamma = \frac{(\sigma \Omega_0)^2}{2R_{\rm s}}
\end{equation}

In our experiment, the systems decoheres for durations of $5 \mu s, 10 \mu s$ and $50 \mu s$ and the noise sampling rate is set to $1\,\mathrm{MHz}$. Hence the decoherence rate is given by $\gamma_{\rm exp} = (\sigma \Omega_0)^2/2\,\mathrm{MHz}$ when $\Omega_0$ is measured in $\mathrm{MHz}$.
\\

\begin{figure}[ht]\centering
\includegraphics[width=0.7\textwidth]{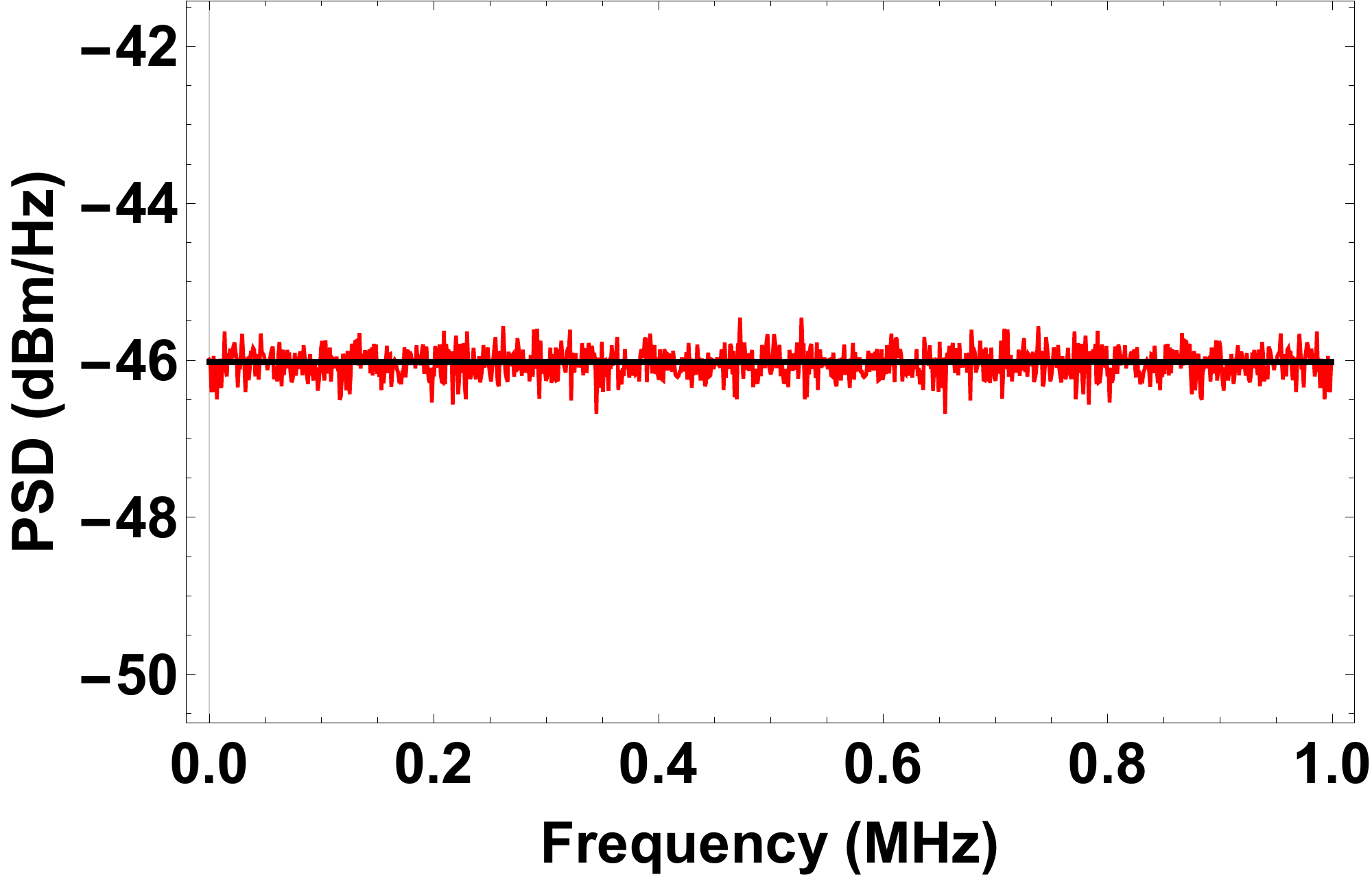}
\caption{Power spectral density of discrete Gaussian white noise with standard deviation $\sigma = 5$ and sampling rate $1\,\mathrm{MHz}$.}
\end{figure}

\begin{figure}[ht]\centering
\includegraphics[width=0.7\textwidth]{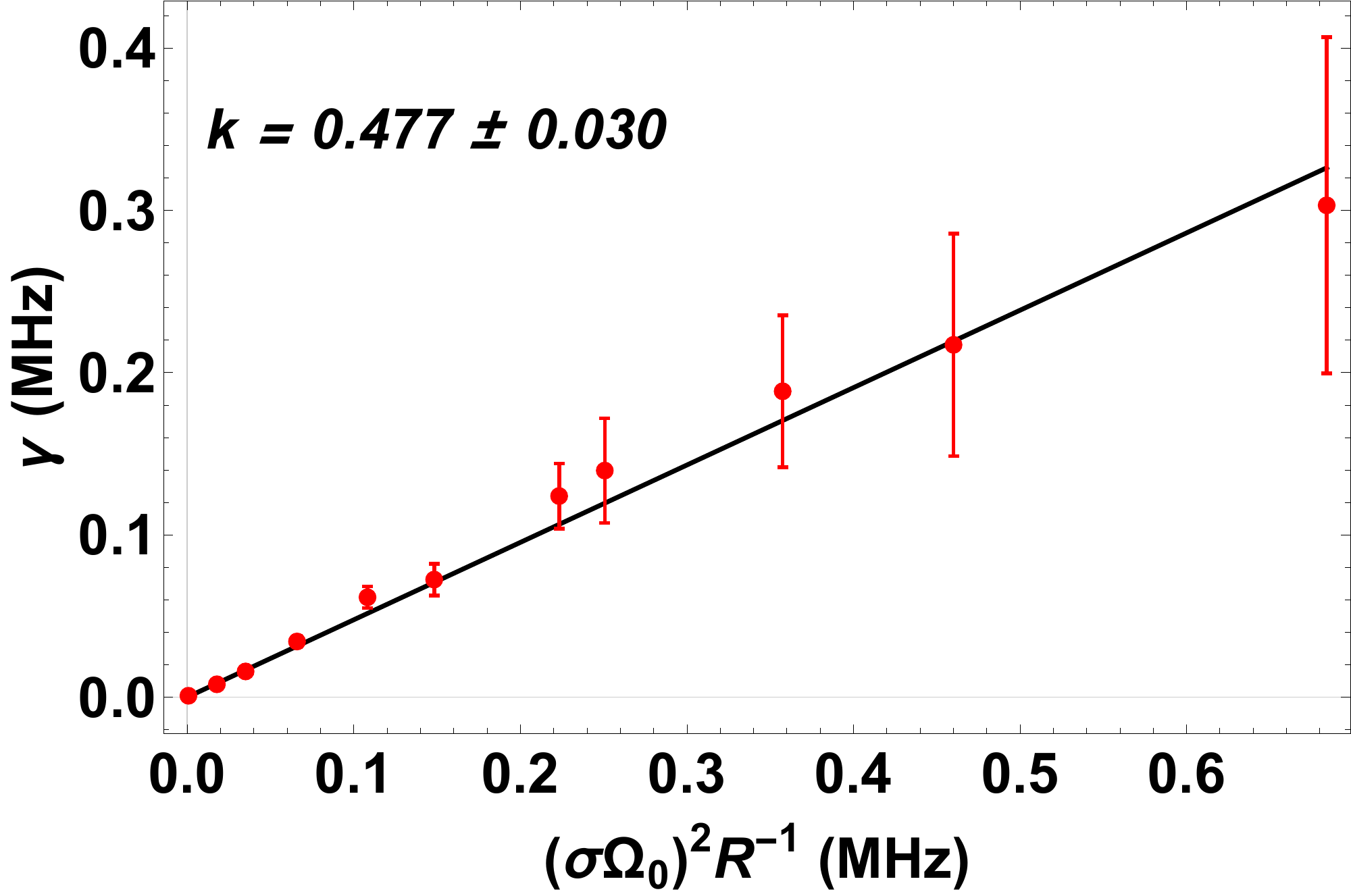}
\caption{Experiment results of decoherence rate $\gamma$ relation with $(\sigma\Omega_0)^2 R_{\rm s}^{-1}$. Here sampling rate is set as $1\,\mathrm{MHz}$.}
\end{figure}

\section{Thermal State Preparation}
\begin{figure}[ht]\centering
\includegraphics[width=0.7\textwidth]{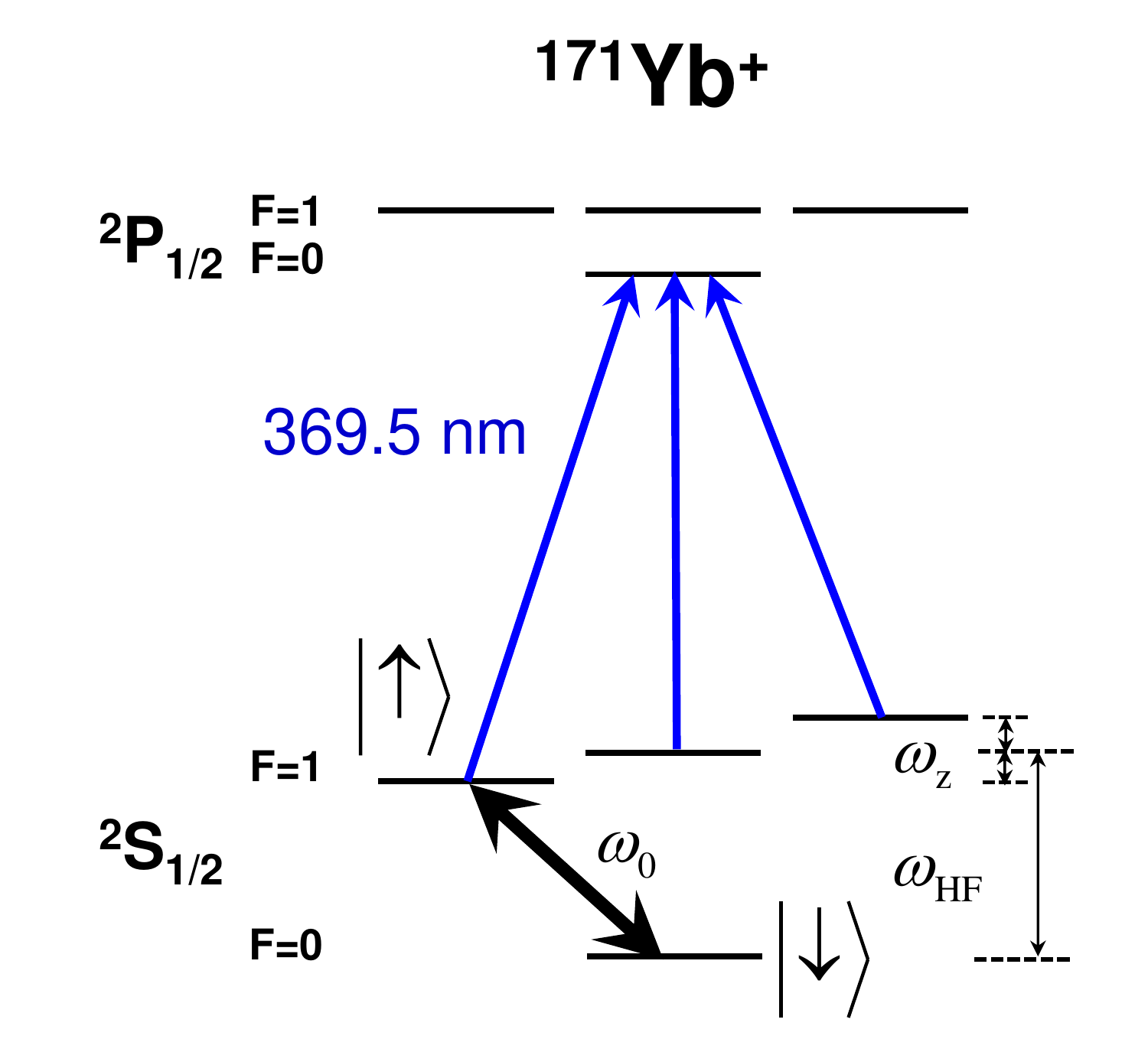}
\caption{Energy levels of our \Yb ion system.  The two level system used in our experiment is composed from the states $\ket{\uparrow}$  and $\ket{\downarrow}$.  Transitions between these states are driven using resonant microwaves.}\label{fig:Yb}
\end{figure}

\begin{figure}[ht]\centering
\includegraphics[width=0.7\textwidth]{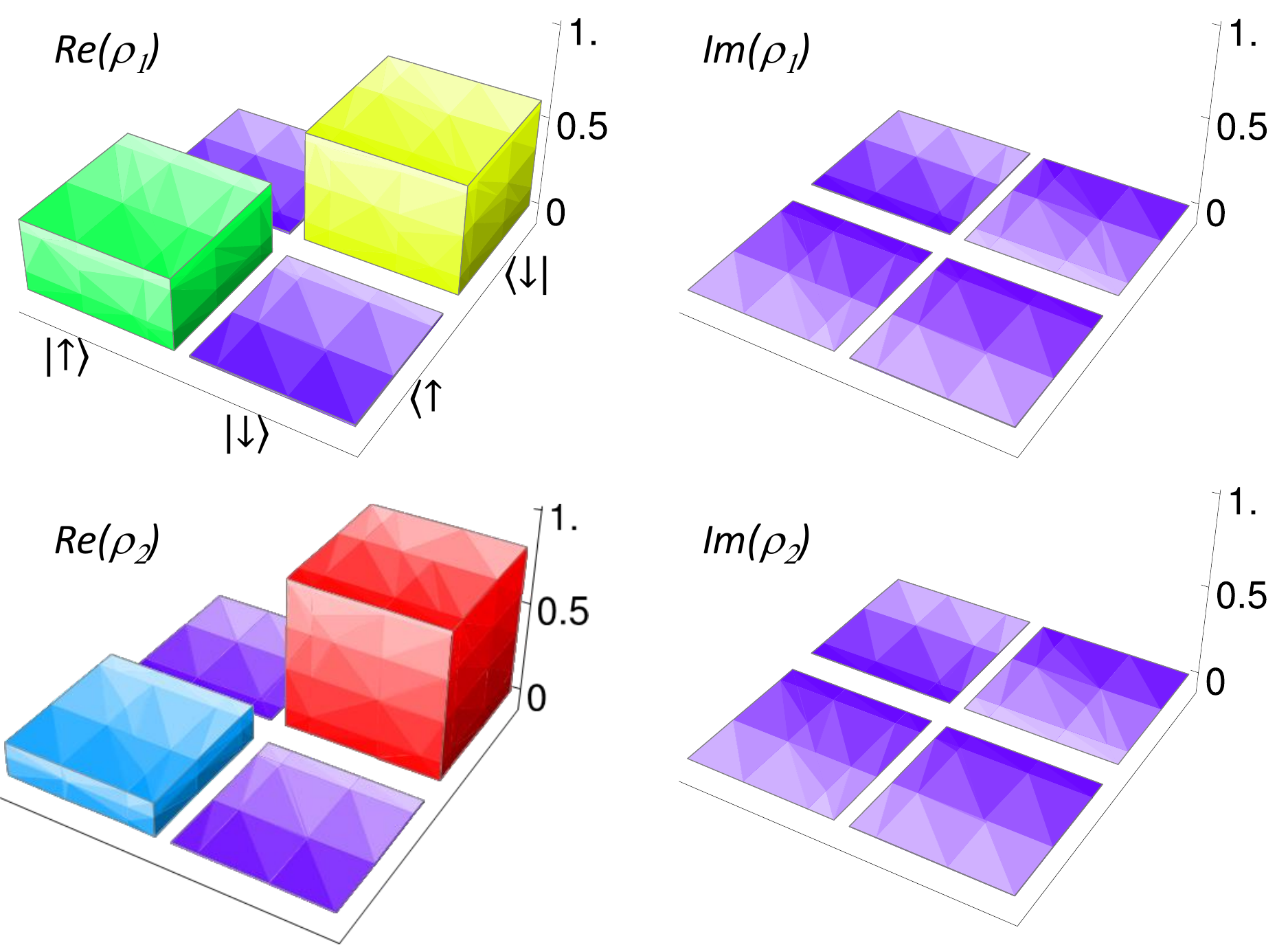}
\caption{Density matrices after preparing effective thermal states, which are equivalent to (a) $T^{eq}_1=5.63\;\mu$K and (b) $T^{eq}_2=1.70\;\mu$K.}\label{fig:Thermalization}
\end{figure}

We use the magnetic field sensitive states $|^2S_{1/2},F=1,m_F=-1\rangle \equiv \ket{\uparrow}$ and $|^2S_{1/2},F=0,m_F=0\rangle \equiv \ket{\downarrow}$ to create an effective two state system with a typical coherence time of 0.14 ms. After preparing a superposition state with the desired populations, we wait 1.5 ms for the system to decohere. We confirm that the state is effectively thermal using state-tomography \cite{Xiang13}. As shown in Fig. \ref{fig:Thermalization}, the off-diagonal components of the density matrix are negligible for both effective temperatures used in our setup. 

\subsection{Energy Measurements}
The first and the second energy measurements are performed in the $\hat{\sigma}_{\rm z}$ basis using standard fluorescence detection as shown in Fig. \ref{fig:Yb}. Depending on whether the system is in the excited state $\ket{\uparrow}$ or ground state $\ket{\downarrow}$, fluorescence or no fluorescence  respectively occurs during the detection sequence. When the ground state $\ket{\downarrow}$ (dark state) is measured, the system remains unchanged during the detection sequence and we simply continue to the next step of the experiment.  If the excited state $\ket{\uparrow}$ (bright state) is detected, the system is left in a mixture of the three levels of $F=1$ in $^2$S$_{1/2}$ manifold. Therefore, we re-prepare the $\ket{\uparrow}$ state using standard optical pumping and a $\pi$-pulse of microwaves before continuing the experiment.  A fluorescence detection sequence is also used for the final measurement which constitutes the end of an experimental run.

\subsection{Adiabatic Rotation}
For our setup, the initial and the final energy measurements are performed in the $\hat{\sigma}_z$ basis. Between the measurement sequences and the driving protocol, the state of the system must be transferred between the $z$-axis and $x$-$y$ plane of the Bloch sphere. To accomplish this task, we use adiabatic shortcuts -- a protocol that has the same effect as an adiabatic switching of the Hamiltonian but occurs in finite time \cite{Rice03,Berry09,An16}.  Specifically we apply an additional counterdiabatic term to our Hamiltonian during the switching process to achieve the shortcut.

 After thermal state preparation and the first energy measurement, our system collapses into the $\ket{\uparrow}$ or $\ket{\downarrow}$ state.  In principle, we have to adiabatically rotate the $\ket{\uparrow}$ or $\ket{\downarrow}$ state to the corresponding state in the x-y plane of the Bloch sphere. In our experiment, the coherence time of a superposition of the $\ket{\uparrow}$ and $\ket{\downarrow}$ states is short and hence would introduce an error in the rotation if it were carried out in a truly adiabatic fashion. Therefore, we apply an adiabatic shortcut to reduce the time for the rotation. In this scheme, we change the Hamiltonian of the system according to
\begin{equation}
\hat{H_{1}}(t)=\frac{\Delta_{0}}{2}\hat\sigma_{\rm z}\cos{(\omega_{1} t)}+\frac{\Omega_{0}}{2}(\hat\sigma_{\rm x}\sin{(\omega_{1} t)}+\hat\sigma_{\rm y})
\end{equation}
where $\omega_1=\Omega_{0}=\Delta_{0}=(2\pi) 50$ kHz and $t$ varies from $t=0$ to $t =\pi/2\omega_1=5~\mu$s.   The term proportional to $\hat\sigma_{\rm y}$ is the counterdiabatic which suppresses the excitations. Note that true adiabatic rotation requires at least hundreds of $\mu$s, which is much longer than transfer time using the adiabatic shortcut. 

After the driving sequence, we rotate the system's state back to the z-axis of the Bloch sphere using the Hamiltonian 
\begin{equation}
\hat{H_{2}}(t)=\frac{\Omega_{0}}{4}(\hat\sigma_{\rm y}\cos{(\omega_{2} t)}+\hat\sigma_{\rm x})+ \frac{\Delta_{0}}{4}\hat\sigma_{\rm z}\sin{(\omega_{2} t)}
\end{equation}
where $\omega_2=\Omega_{0}/2=\Delta_{0}/2=(2\pi) 25$ kHz and $t$ varies from $t=0$ to $t =\pi/2\omega_2=10~\mu$s. This time the courterdiabatic term is proportional to  $\hat\sigma_{\rm x}$.

\section*{Acknowledgements.} 
The authors thank Janet Anders, Alexia Auff{\` e}ves, Sebastian Deffner, and Ken Wright for their discussion and comments pertaining to this manuscript. This work was supported by the National Key Research and Development Program of China under Grants No. 2016YFA0301900 (No. 2016YFA0301901), the National Natural Science Foundation of China 11374178, 11574002 and 11504197. AS and CJ were supported by the United States National Science foundation under grant DMR-1506969. HTQ also acknowledges support from the National Science Foundation of China under grants 11375012 and 11534002.\\

\section*{References}


\end{document}